\documentclass[aps,pra,twocolumn,floatfix,groupedaddress,superscriptaddress,footinbib,notitlepage,showpacs]{revtex4-2}
\usepackage{times,bm,bbm,bbold,amssymb,amsbsy,amsthm,amsmath,amsfonts,graphicx,graphics,color,xcolor,hyperref}
\hypersetup{colorlinks,linkcolor={blue},citecolor={blue},urlcolor={blue}}
\newcommand{\ignore}[1]{}
\DeclareFontFamily{OT1}{pzc}{}
\DeclareFontShape{OT1}{pzc}{m}{it}
              {<-> s * [1.25] pzcmi7t}{}
\DeclareMathAlphabet{\mathpzc}{OT1}{pzc}
                                 {m}{it}
\usepackage[T1]{fontenc}
\usepackage{newtxtext,newtxmath}

\begin{document}

\title{Correlation-Enabled Energy Exchange in Quantum Systems without External Driving}

\author{T. Pyh\"{a}ranta}
\affiliation{QTF Center of Excellence, Department of Applied Physics, Aalto University, P. O. Box 11000, FI-00076 Aalto, Espoo, Finland}

\author{S. Alipour}
\email{sahar.alipour@aalto.fi}
\affiliation{QTF Center of Excellence, Department of Applied Physics, Aalto University, P. O. Box 11000, FI-00076 Aalto, Espoo, Finland}

\author{A. T. Rezakhani}
\affiliation{Department of Physics, Sharif University of Technology, Tehran 14588, Iran}
\affiliation{School of Physics, Institute for Research in Fundamental Sciences (IPM), Tehran 19538, Iran}

\author{T. Ala-Nissila}
\affiliation{QTF Center of Excellence, Department of Applied Physics, Aalto University, P. O. Box 11000, FI-00076 Aalto, Espoo, Finland}
\affiliation{Interdisciplinary Center of Mathematical Modelling and Department of Mathematical Sciences, Loughborough University, Loughborough, Leicestershire LE11 3TU, UK}

\begin{abstract}
\noindent We study the role of correlation in mechanisms of energy exchange between an interacting bipartite quantum system and its environment by decomposing the energy of the system to \textit{local} and \textit{correlation-related} contributions. When the system Hamiltonian is time-independent, no external work is performed. In this case, energy exchange between the system and its environment occurs only due to the change in the state of the system. We investigate possibility of a special case where the energy exchange with the environment occurs exclusively due to changes in the correlation between the constituent parts of the bipartite system, while their local energies remain constant. 
We find sufficient conditions for preserving local energies. It is proven that under these conditions and within the Gorini-Kossakowski-Lindblad-Sudarshan (GKLS) dynamics this scenario is not possible for \textit{all} initial states of the bipartite system. Nevertheless, it is still possible to find special initial states for which the local energies remain unchanged during the associated evolution and the whole energy exchange is only due to the change in the correlation energy. We illustrate our results with an example.  
\end{abstract}
\date{\today}
\maketitle

\textit{Introduction.---}Quantum thermodynamics is a nascent branch of physics concerned with understanding the thermodynamic behavior of quantum systems and generalizing the laws of thermodynamics to quantum systems to account for the impact of inherently quantum mechanical phenomena. 
Correlation is one of the main features in composite quantum systems and is considered as a resource for various applications. Within the framework of thermodynamics, it is possible to study different correlations such as system-environment correlation, correlation between different constituents of the system itself, or even correlations within the environment. It is known that system-environment correlation is an important element in the system's dynamical equation and accordingly in the thermodynamic properties of the system, e.g., in energy transfer \cite{Pekola-system-bath-corr-heat, Lloyd-noenergy-discord, SciRep:Alipour-corr,Sampaio-ConditionalWaveF}. In particular, in the strong coupling regime correlation comes into play indirectly through interactions \cite{Rivas-nonEquilibMeanForce,Miller-Anders-MeanForce}. A direct approach for explicitly considering correlation has been introduced in Ref. \cite{ULL,unfolding-corr}. The effect of correlation in the thermodynamic arrow of time has been studied in literature and it has been shown that correlation can lead to reversal of the direction of heat flow from cold to hot \cite{Partovi-ThermoTimeArrow,Rudolph-ThermoTimeArrow, Lutz-reversing-HeatFlow}. In Ref. \cite{Ptazynski-Esposito-IntraEnvironmentCorr}, the effect of intra-environmental correlation in open-system entropy production has been studied. 

The second case, i.e., the role of inside-the-system or {\it intra-system} correlations has been explored in different contexts such as binding energy \cite{SciRep:Alipour-corr,binding}, latent heat \cite{Morawetz-corr-latent-heat,Vedral-latent-work}, locality of temperature \cite{Gogolin-LocalityTemp}, and in the relation between the temperatures of the two parts of a bipartite system \cite{qtemperature}. Along these lines, much recent attention has been paid on the energy considerations of creating and destroying correlations in a bipartite system. Most of such studies involve unitary evolution of closed bipartite systems with driven Hamiltonians. Within this framework, it has been shown that it is possible to extract work from correlations by obtaining bounds on the extractable work from globally correlated but locally thermal subsystems  \cite{Brandao-extractable-work,Alicki-extractable-work,Huber-Extractable-work-corr,Giovanetti-Extractable-work-corr,Manzano-work-corr,delCampo-fluctuation-battery}. The work cost of creating correlations in a system with initially uncorrelated thermal subsystems has also been studied in Refs. \cite{Rudolph-MaxMinCorrWithUnitary,Bakhshinezhad-optimal-creation-corr}. 

In this work, we focus on the effect of correlations between the parts of an interacting bipartite system in energy exchange with the environment. 
The main question of interest is whether it is possible to exchange energy between the bipartite system and its environment such that the only effect the process has on the system is to change the correlation between the two parts of the system. To this end, to avoid energy exchanges due to driving the system Hamiltonian (or, equivalently, due to external work applied on or performed by the system) we consider cases where the system Hamiltonian is time-independent. In such cases, the whole energy change in the system is environment-induced, because there is no external agent. Since the internal energy of the system is defined as the expectation value of the system Hamiltonian, the whole energy change in the system is due to the change in the state of the bipartite system when the Hamiltonian is constant. 

To investigate the role of correlation in energy exchange, we need to clearly identify the contribution of correlation to the internal energy of the system \cite{SciRep:Alipour-corr}. To this end, we use a decomposition of the system energy in terms of local energy and correlation energy contributions. In a bipartite system, the correlation energy is part of the internal energy which is locally inaccessible and is defined as the difference between the internal energy of the system and the internal energy assigned to the uncorrelated counterparts of the system. Thus, internal energy is divided into three parts: two parts related to the internal energy of the subsystems and one part related to the correlation between them. This implies that a spatially bipartite system can behave energetically as tripartite, where the correlation energy can be exchanged between the system and the environment independently of the local energy of the subsystems. 

We use the energy decomposition to study the conditions under which dynamics can be universally (i.e., for any initial state) local-energy preserving but not correlation-energy preserving to allow energy exchange through the correlation only. We prove that within the standard Gorini-Kossakowski-Lindblad-Sudarshan (GKLS) dynamics universal local-energy preservation leads in general to correlation-energy preservation (for universal extensions of the GKLS dynamics see Ref. \cite{ULL}). However, we also show that it is possible to find special initial states that allow dynamics to preserve local subsystem energies while correlation-related energy is exchanged between the system and the environment. This facilitates the use of intrasystem correlation as a resource for energy with proper preparation of the bipartite system. We demonstrate this through an example where the subsystems' temperatures and local energies are conserved while the global energy of the system changes due to the change in correlation.
\begin{figure}
\includegraphics[width=\linewidth]{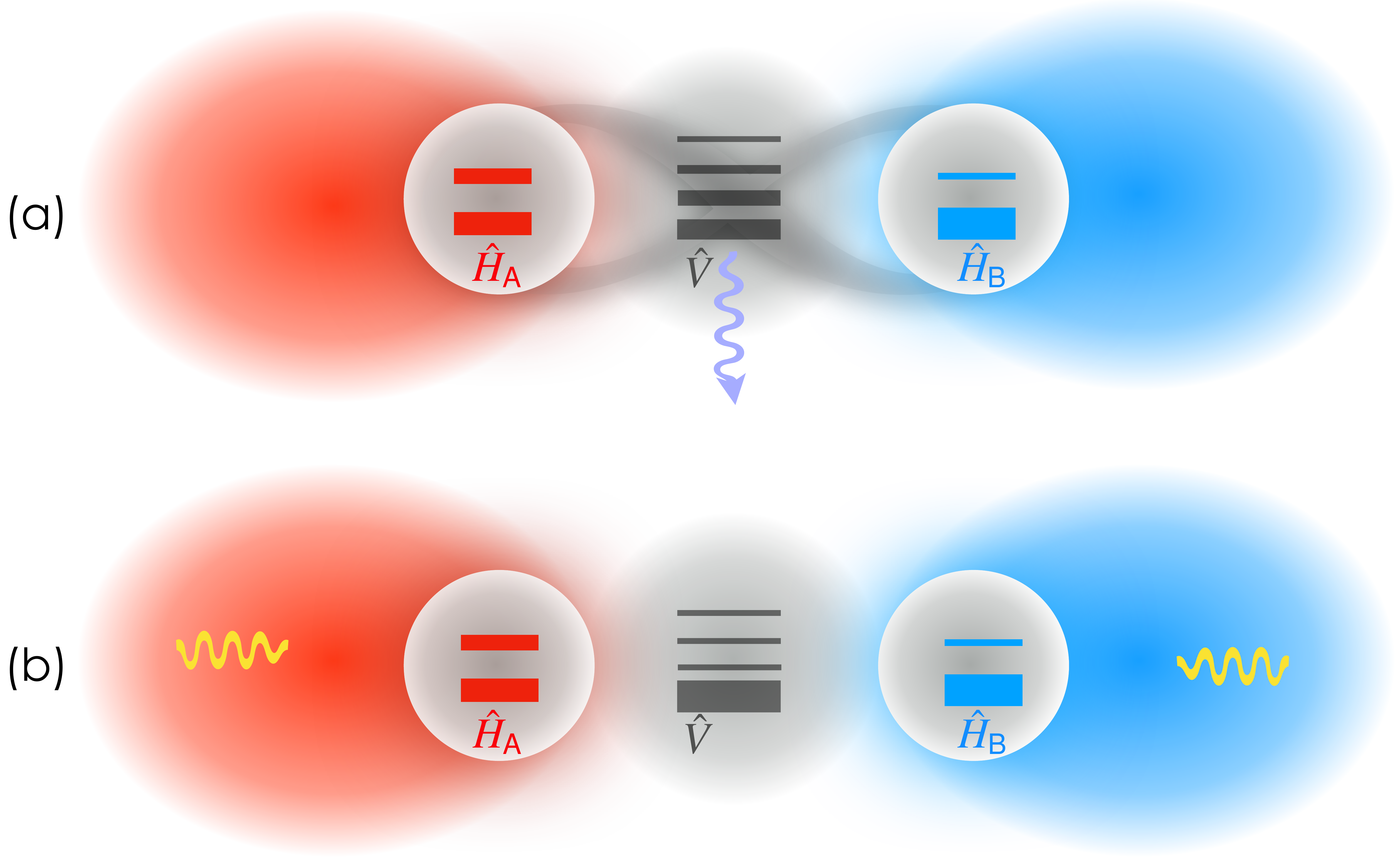}
\caption{Schematic of releasing energy by changing correlation. In addition to the two subsystems, correlation functions as the third part in the energy-exchange mechanisms in the subsystems. (a) The two subsystems $\mathsf{A}$ and $\mathsf{B}$, with Hamiltonians $H_{\mathsf{A}}$ and $H_{\mathsf{B}}$, interact through a Hamiltonian $V$. The energy levels of each Hamiltonian are shown. The subsystems $\mathsf{A}$ and $\mathsf{B}$ are initially at equilibrium with a hot (red) bath and cold (blue) bath, respectively. The gray chain connecting the subsystems represents correlation. The thickness of an energy level represents the population of that level. The purple arrow shows a transition in the $V$ levels, which is only related to the change in correlation (this can be seen using the relation $\mathrm{Tr}[\varrho \hat{V}]=\mathrm{Tr}[\chi \hat{V}]=\mathrm{Tr}[\chi V]$). (b) Transition to the lower level of $V$ is accompanied by correlation removal and releasing energy into the bath (shown in yellow) without any transitions in the local energy levels.}
\label{fig:Correlations}
\end{figure}

\textit{The model.---}We consider an interacting bipartite quantum system with density matrix $\varrho$ and time-independent total Hamiltonian $H = H_{\mathsf{A}} + H_{\mathsf{B}} + V$, comprised of the bare subsystem Hamiltonians $H_{\mathsf{A,B}}$ and an interaction term $V$. Each subsystem is weakly coupled with a separate heat bath such that the nonunitary effects of the baths on the system dynamics are described by local Markovian Lindblad superoperators $\mathpzc{L}_{\mathsf{A}}$ and $\mathpzc{L}_{\mathsf{B}}$ \cite{Lindblad76, Gorini76}, and the system dynamics is described by the summation of the local superoperators as the following (for a discussion on adding dynamical generators see Ref. \cite{adding-GKLSgenerators}):
\begin{align}
\dot{\varrho} = - i[H,\varrho] + \mathpzc{L}_{\mathsf{A}}[\varrho] + \mathpzc{L}_{\mathsf{B}}[\varrho], 
\label{eq:LindbladEquation}
\end{align}
where we have set $\hbar =1$ and dot denotes time derivative. Here, 
\begin{align}
\mathpzc{L}_{\mathsf{A,B}}[\circ] &\equiv \textstyle{\sum_{\mu}}\big( L_\mu^\mathsf{A,B} \circ L_\mu^\mathsf{A,B\dagger} - (1/2) \{\circ, L_\mu^\mathsf{A,B\dagger} L_\mu^\mathsf{A,B}\}\big),  
\end{align}
with $L_\mu^\mathsf{A,B}$ denoting the (generally non-Hermitian) jump operators and $\{A,B\} \equiv AB + BA$ the anticommutator, and we have neglected the Lamb-shift corrections. 

To obtain the dynamics of each subsystem, we note that the system density matrix can be decomposed as 
\begin{align}
\varrho = \varrho_{\mathsf{A}} \otimes \varrho_{\mathsf{B}} + \chi, 
\label{state-decompos-chi}
\end{align}
where the correlated part $\chi$, satisfying the partial-trace condition $\mathrm{Tr}_{\mathsf{A,B}}[\chi] = 0$ with respect to either subsystem, is the difference between the state of the total system $\varrho$ and the tensor product of the subsystem states $\varrho_{\mathsf{A,B}} \equiv \mathrm{Tr}_{\mathsf{B,A}}[\varrho]$. Substituting this decomposition into Eq. \eqref{eq:LindbladEquation} and tracing out the excess degrees of freedom yields  \cite{SciRep:Alipour-corr} 
\begin{align}
\dot{\varrho}_{\mathsf{A,B}} = - i[\hat{H}_{\mathsf{A,B}}, \varrho_{\mathsf{A,B}}] + \mathpzc{L}_{\mathsf{A,B}}[\varrho_{\mathsf{A,B}}] -  i \, \mathrm{Tr}_{\mathsf{B,A}}\big[[V,\chi]\big], 
\label{eq:SubsystemLindbladEquations}
\end{align}
where
\begin{align}
\hat{H}_{\mathsf{A,B}} \equiv  H_{\mathsf{A,B}} + \mathrm{Tr}_{\mathsf{B,A}}[V \varrho_{\mathsf{B,A}}] 
- \alpha_\mathsf{A,B} \, \mathrm{Tr}[V \varrho_{\mathsf{A}} \otimes \varrho_{\mathsf{B}}] \mathbbmss{I}, 
\label{eq:EffectiveHamiltonians}
\end{align}
are effective local Hamiltonians, with $\alpha_{\mathsf{A,B}}$ constants satisfying $\alpha_{\mathsf{A}} + \alpha_{\mathsf{B}} = 1$. The total Hamiltonian can be recast in terms of the effective Hamiltonians as 
\begin{align}
H = \hat{H}_{\mathsf{A}} + \hat{H}_{\mathsf{B}} + \hat{V},
\label{H-reshaped}
\end{align}
where $\hat{V} \equiv  V - \mathrm{Tr}_{\mathsf{A}}[ V \varrho_{\mathsf{A}}] - \mathrm{Tr}_{\mathsf{B}}[ V \varrho_{\mathsf{B}}]
+ \mathrm{Tr}[V \varrho_{\mathsf{A}} \otimes \varrho_{\mathsf{B}}] \mathbbmss{I}$ is the effective interaction Hamiltonian. It should be noted that the effective local and interaction Hamiltonians are time-dependent due to their state dependence. 

\textit{Energy decomposition}.---Using the state decomposition \eqref{state-decompos-chi}, it can be seen that the total energy $U=\mathrm{Tr}[\varrho H]$ of the system contains local-energy $U_{\otimes}$ and correlation-energy $U_{\chi}$ contributions. The local part is given by
\begin{align}
 U_{\otimes}\equiv\mathrm{Tr}[\varrho_{\mathsf{A}}\otimes \varrho_{\mathsf{B}} H] = \mathrm{Tr}[\varrho_{\mathsf{A}} \hat{H}_{\mathsf{A}}]+\mathrm{Tr}[\varrho_{\mathsf{B}} \hat{H}_{\mathsf{B}}]\equiv U_{\mathsf{A}}+U_{\mathsf{B}},
 \end{align}
where $U_{\mathsf{A,B}}$ are the internal energies assigned to the subsystems. The local-energy part can be obtained by locally measuring the effective Hamiltonians. The correlation energy is then naturally given by $U_{\chi}=U-U_{\otimes}=\mathrm{Tr}[(\varrho-\varrho_{\mathsf{A}}\otimes \varrho_{\mathsf{B}}) H]$, which becomes
\begin{align}
U_{\chi}=\mathrm{Tr}[\chi H] = \mathrm{Tr}[\chi V],
\end{align}
where in derivation of the last term we have used the vanishing partial trace properties of $\chi$. 
The correlation energy depends on the correlation operator and the interaction Hamiltonian and is independent of the local Hamiltonians and states. 
This part of the energy is not accessible by the local subsystems and could only be measured globally on the total system. We observe that in the sense of energy, correlation thus behaves as an independent entity in addition to the subsystems---Fig. \ref{fig:Correlations}.

Since $H$ is time-independent, the total energy variation of the system is only due to the energy exchange with the environment through the change in the state of the system as $dU=\mathrm{Tr}[d\varrho \, H]=dU_{\otimes}+dU_{\chi}$. This is related to change in local states leading to local energy change as 
\begin{align}
d U_{\otimes}=\mathrm{Tr}[d(\varrho_{\mathsf{A}}\otimes \varrho_{\mathsf{B}}) H] = \mathrm{Tr}[d\varrho_{\mathsf{A}} \hat{H}_{\mathsf{A}}]+\mathrm{Tr}[d\varrho_{\mathsf{B}} \hat{H}_{\mathsf{B}}],
\label{local-energy-change}
\end{align}
and change in the correlated part of the state, which is related to change in the correlation energy as 
\begin{align}
d U_{\chi}=\mathrm{Tr}[V\,d\chi].
\label{correlation-energy-change}
\end{align}

In the following, we investigate whether it is possible to find a set of microscopic conditions for the dynamics under which \textit{for any given initial state} the local energy is conserved while the correlation energy is not, so that the whole system energy exchange with the environment is only through correlation. 

\textit{Conditions for local energy conservation.---}To explore whether open-system dynamics given by Eq. \eqref{eq:LindbladEquation} can be in general local-energy preserving, we modify the condition $dU_{\otimes}=0$ by inserting $d\varrho_{\mathsf{A},\mathsf{B}}$ from Eq. \eqref{eq:SubsystemLindbladEquations} into $dU_{\otimes}$ of Eq. \eqref{local-energy-change}. We then use the cyclic property of trace to obtain $\mathrm{Tr}\big[ \hat{H}_{\mathsf{A,B}} [V, \chi] \big] = \mathrm{Tr}\big[ [\hat{H}_{\mathsf{A,B}}, V] \chi \big]$ and $\mathrm{Tr}_{\mathsf{A,B}}\big[ \hat{H}_{\mathsf{A,B}} \mathpzc{L}_{\mathsf{A,B}}[\varrho_{\mathsf{A,B}}] \big] = \mathrm{Tr}_{\mathsf{A,B}}\big[ \mathpzc{L}_{\mathsf{A,B}}^{\#}[\hat{H}_{\mathsf{A,B}}] \varrho_{\mathsf{A,B}} \big]$, where the Hilbert-Schmidt adjoints $\mathpzc{L}_{\mathsf{A,B}}^{\#}$ are defined by 
\begin{align}
\mathpzc{L}_{\mathsf{A,B}}^{\#}[\circ] &\equiv \textstyle{\sum_{\mu}} L_\mu^\mathsf{A,B\dagger} \circ L_\mu^\mathsf{A,B} - (1/2) \{\circ, L_\mu^\mathsf{A,B\dagger} L_\mu^\mathsf{A,B}\}. 
\end{align}
It can be seen that $d U_{\otimes}$ can vanish regardless of the instantaneous state of the bipartite system when the following sufficient conditions are met (Appendix \ref{App}): (i) For an arbitrary instantaneous state, we must have $[\hat{H}_{\mathsf{A}}+\hat{H}_{\mathsf{B}},V] = 0$ at all times. For example, if $V=O_{\mathsf{A}}\otimes O_{\mathsf{B}}$, for some observables $O_{\mathsf{A}}$ and $O_{\mathsf{B}}$, such that $[H_{\mathsf{A}}+H_{\mathsf{B}}, O_{\mathsf{A}}\otimes O_{\mathsf{B}}]=0$, it can be seen that condition (i) is satisfied independently of the state of the bipartite system. 
(ii) We must also have $\mathpzc{L}_{\mathsf{A}}^{\#}[H]+\mathpzc{L}_{\mathsf{B}}^{\#}[H]= 0$.
Condition (ii) implies that local energy conservation generally leads to total energy conservation, which gives a negative answer to our main question about having energy exchange with the environment through the correlation energy only. This can be seen by noting that $\mathrm{Tr}\big[(\mathpzc{L}_{\mathsf{A}}^{\#}[H]+\mathpzc{L}_{\mathsf{B}}^{\#}[H])\varrho \big]=\mathrm{Tr}\big[-i([H,\varrho]+\mathpzc{L}_{\mathsf{A}}[\varrho]+\mathpzc{L}_{\mathsf{B}}[\varrho])H \big]= \mathrm{Tr}\big[d\varrho\, H \big]=dU=0$. It should be noted that this result is not an effect of local environments and that the proof can be straightforwardly extended to the case the environment acts globally of the subsystems (not in a local fashion). This leads to the following theorem:

\textbf{Theorem:} Local energy conservation in an open bipartite system with a GKLS dynamics, satisfying the two conditions above, leads to the conservation of the total energy, i.e., no energy can be exchanged between the system and the environment. 

However, we show below, through an example, that although dynamics in general cannot exchange the correlation energy while it preserves the local energy, there exist special cases of the system states for which it is possible to exchange energy with the environment while local energy is still preserved. 

\textit{Example: Thermalizing dynamics.---}We assume that both parts of the system weakly interact with separate heat baths at different inverse temperatures $\beta_{\mathsf{A}}$ and $\beta_{\mathsf{B}}$. The system dynamics is given by Eq. \eqref{eq:LindbladEquation}, and the jump operators describe transitions between different eigenstates of the corresponding bare subsystem Hamiltonians $L_{mn}^\mathsf{A} \equiv |m\rangle_{\mathsf{A}}\langle n|$ and $L_{mn}^\mathsf{B} \equiv |m\rangle_{\mathsf{B}}\langle n|$, where $|m\rangle_{\mathsf{A,B}}$ and $|n\rangle_{\mathsf{A,B}}$ are different eigenvectors of the bare local Hamiltonians $H_{\mathsf{A,B}}$. The dissipative parts of the dynamical equation are given by 
\begin{align}
&\mathpzc{L}_{\mathsf{A,B}}[\circ]=\textstyle{\sum_{m \neq n}} \gamma_{mn}^\mathsf{A,B} \big( L_{mn}^\mathsf{A,B} \circ L_{mn}^\mathsf{A,B\dagger} - (1/2) \{L_{mn}^\mathsf{A,B\dagger} L_{mn}^\mathsf{A,B}, \circ \} \big), 
\label{eq:DissipatorExample}
\end{align}
where $\gamma_{mn}^\mathsf{A,B}$ are jump rates. 
We consider here the case  where the dissipators have thermal steady states \cite{Ostilli-thermalization}. 
To this end, we assume that the transition rates satisfy the detailed-balance condition 
\begin{align}
\gamma_{mn}^\mathsf{A,B} = \gamma_{nm}^\mathsf{A,B}\, e^{-\beta_{\mathsf{A,B}}(E_m^\mathsf{A,B}-E_n^\mathsf{A,B})}, \label{eq:DetailedBalanceCondition}
\end{align}
where we have denoted the eigenvalues of the bare local Hamiltonians by $E_n^\mathsf{A,B}$. Assuming that $H_{\mathsf{A,B}}$ have nondegenerate spectra, this guarantees that $\pi_{\mathsf{A}} \equiv  e^{-\beta_{\mathsf{A}} H_{\mathsf{A}}} / Z_{\mathsf{A}}$ and $\pi_{\mathsf{B}} \equiv e^{-\beta_{\mathsf{B}} H_{\mathsf{B}}} / Z_{\mathsf{B}}$ are the unique steady states of $\mathpzc{L}_{\mathsf{A}}$ and $\mathpzc{L}_{\mathsf{B}}$, respectively, where  $Z_{\mathsf{A,B}} \equiv \mathrm{Tr}_{\mathsf{A,B}}[ e^{-\beta_{\mathsf{A,B}}H_{\mathsf{A,B}}}]$. By imposing the commutation condition $[H_{\mathsf{A}}+H_{\mathsf{B}}, V]=0$, we get that
\begin{align}
\pi_{\mathsf{A}} \otimes \pi_{\mathsf{B}} =  e^{-\beta_{\mathsf{A}} H_{\mathsf{A}}} / Z_{\mathsf{A}} \otimes e^{-\beta_{\mathsf{B}} H_{\mathsf{B}}} / Z_{\mathsf{B}}
\end{align}
is the unique steady state of the dynamics \eqref{eq:LindbladEquation}. Let us now assume that the system is initially prepared in the state $\varrho(0)= \pi_{\mathsf{A}} \otimes \pi_{\mathsf{B}}+\chi(0)$. Bearing in mind that $\pi_{\mathsf{A}} \otimes \pi_{\mathsf{B}}$ is the steady state of the dynamics \eqref{eq:LindbladEquation}, only the correlation part of the total density operator $\varrho(t) = \pi_{\mathsf{A}} \otimes \pi_{\mathsf{B}} + \chi(t)$ evolves in time. This leads to a change in the total energy, while the local energy remains conserved. Accordingly, the energy exchanged is fully determined by the changes in the nonlocal contributions to the internal energy, given simply by
\begin{align}
\begin{split}
\Delta U_{\chi} (t) =& \mathrm{Tr}[V\big(\chi(t) - \chi(0)\big)]. \label{eq:QubitCorrelatedHeat}
\end{split}
\end{align}

As a simple demonstration we consider two qubits with bare Hamiltonians $H_{\mathsf{A}}=\omega_{\mathsf{A}} \sigma_z\otimes \mathbbmss{I}_{\mathsf{B}}$ and $H_{\mathsf{B}}=\omega_{\mathsf{B}} \mathbbmss{I}_{\mathsf{A}}\otimes\sigma_z$, where $\omega_{\mathsf{A,B}}\geqslant 0$, coupled through an interaction 
$V = g \sigma_{z}\otimes \sigma_{z}$, where $g$ is the coupling constant and $\sigma_{z}=|0\rangle\langle0| - |1\rangle\langle1|$ is the Pauli operator in the $z$ direction. Assume that $L_{10}^{\mathsf{A}}=|1\rangle\langle 0|\otimes \mathbbmss{I}_{\mathsf{B}}$, $L_{01}^{\mathsf{A}}=|0\rangle\langle 1| \otimes \mathbbmss{I}_{\mathsf{B}}$, $\gamma_{01}^{\mathsf{A}}=e^{-\beta_{\mathsf{A}} \omega_{\mathsf{A}}}$, and $\gamma_{10}^{\mathsf{A}}=e^{\beta_{\mathsf{A}} \omega_{\mathsf{A}} }$. 
Similarly, $L_{10}^{\mathsf{B}}=\mathbbmss{I}_{\mathsf{A}} \otimes |1\rangle\langle 0|$, $L_{01}^{\mathsf{B}}=\mathbbmss{I}_{\mathsf{A}} \otimes |0\rangle\langle 1|$, $\gamma_{01}^{\mathsf{B}}=e^{-\beta_{\mathsf{B}} \omega_{\mathsf{B}}}$, and $\gamma_{10}^{\mathsf{B}}=e^{\beta_{\mathsf{B}} \omega_{\mathsf{B}} }$. We prepare the system in a correlated state $\varrho(0)=e^{-\beta_{\mathsf{A}}\omega_{\mathsf{A}}\sigma_z}\otimes e^{-\beta_{\mathsf{B}}\omega_{\mathsf{B}}\sigma_z}/(Z_{\mathsf{A}}Z_{\mathsf{B}})+\chi(0)$, with $Z_{\mathsf{A}}=\mathrm{Tr}[e^{-\beta_{\mathsf{A}}\omega_{\mathsf{A}}\sigma_z}]$, $Z_{\mathsf{B}}=\mathrm{Tr}[e^{-\beta_{\mathsf{B}}\omega_{\mathsf{B}}\sigma_z}]$, and $\chi(0) = c\,\sigma_{z}^\mathsf{A} \otimes \sigma_{z}^\mathsf{B}$, where the range of validity of the parameter $c$ should be consistent with positivity of the total density operator, hence, $c\in [-e^{-\beta_{\mathsf{A}}\omega_{\mathsf{A}}-\beta_{\mathsf{B}}\omega_{\mathsf{B}}}, \min\{e^{-\beta_{\mathsf{A}}\omega_{\mathsf{A}} +\beta_{\mathsf{B}} \omega_{\mathsf{B}}}, e^{\beta_{\mathsf{A}}\omega_{\mathsf{A}}-\beta_{\mathsf{B}}\omega_{\mathsf{B}}}\}]$. Substituting the jump rates and operators into Eq. \eqref{eq:LindbladEquation}, it is straightforward to see that subsystems $\mathsf{A}$ and $\mathsf{B}$ remain intact and in thermal equilibrium with their heat baths, while the correlation operator evolves as 
\begin{align}
\chi(t) = e^{ (\mathpzc{L}_{\mathsf{A}}+\mathpzc{L}_{\mathsf{B}})t}[\chi(0)] =  e^{-\lambda t} \chi(0), 
\end{align}
where $\lambda =\gamma_{01}^\mathsf{A}+\gamma_{10}^\mathsf{A}+\gamma_{01}^\mathsf{B}+\gamma_{10}^\mathsf{B}$. Since $\lambda$ is real and positive, the correlation operator vanishes in the long-time limit: $\chi(\infty)=0$. It is important to note that in this special case both $U_{\mathsf{A}}$ and $U_{\mathsf{B}}$ remain separately constant, whereas during the evolution we get from Eq. \eqref{eq:QubitCorrelatedHeat} that the correlation energy varies as 
\begin{align}
\Delta U_{\chi} (t) = 4 gc (e^{-\lambda t} - 1). 
\end{align}
Since $\mathrm{sign}[\Delta U_{\chi} (t) ]=-\mathrm{sign}[g c]$, if $gc > 0$ the system releases energy to the environment through the change in its correlation operator; and if $gc <0$, the system absorbs energy from the environment. In the asymptotic limit where the system state becomes uncorrelated, we obtain $\Delta U_{\chi} (\infty)=-4gc$, which depending on the sign of $gc$ implies that removing correlation can be both energy consuming or energy producing depending on the microscopic details of the system. 

\textit{Remark}.---According to the conventional definitions of heat and work given by $\delta Q = \mathrm{Tr}[d\varrho\, H]$ and $\delta W = \mathrm{Tr}[ \varrho\,  dH]$ \cite{Alicki, Spohn-EP}, respectively, when the Hamiltonian is time-independent the whole energy exchange is only of the heat type. In the entropy-based definitions \cite{entropic-based-thermo,Ahmadi-refined}, however, heat is assigned to the energy change due to the change in the eigenvalues of the state and work is assigned to the energy change due to the change in the eigenvectors of the state as well as the change in the system Hamiltonian. In the above example, $\Delta U_{\chi}$ is thus heat in the sense of both conventional \cite{Alicki} and entropy-based definitions \cite{entropic-based-thermo} as the Hamiltonian is constant and the eigenvectors of the state of the system remain constant in time.

\textit{Summary and conclusions.---}By decomposing the internal energy of an interacting bipartite system into local-energy and correlation-energy contributions, we have shown that a spatially bipartite system can have three independent contributions to the total energy. To show that the correlation energy can vary independently of the local energy changes, we have studied sufficient conditions under which local energy remains intact during dynamics. We have proven that a GKLS master equation of the bipartite system, satisfying these sufficient conditions, cannot universally (for any arbitrary initial state) preserve the local energy while not preserving the correlation energy. However, we have also demonstrated that  by careful preparation of the initial state of the system, we can direct the evolution such that energy exchange with the environment occurs only through the correlation change. 

The prospect of injecting energy from a quantum system into a heat bath, or vice versa, without increasing the temperature of local subsystems has profound implications. For example, it is possible to store/extract heat in/from a bipartite system without changing the local temperatures of its constituents \cite{qtemperature}. In particular, our results should be relevant for quantum computers, where the control of heat and temperature is a crucial issue for maintaining their performance. In addition, combining our results and previous studies in the literature which indicate the possibility of reverse heat flow due to correlation \cite{Lutz-reversing-HeatFlow} suggests that correlation can in principle be employed as a knob to control the direction of energy flow, e.g., to realize energy transistors \cite{heat-transistor}. Finally, we note that the derivation of our result does not require quantumness of the correlation, hence it raises there is a prospect that a similar effect may also occur in classical stochastic manybody systems. 

\textit{Acknowledgments.---}This work has been supported by the Academy of Finland's Center of Excellence QTF Project 312298 and Sharif University of Technology's Office of Vice President for Research and Technology.

\begin{widetext}
\appendix

\section{Details of the proof of the theorem}
\label{App}

The change in the local energy change is given by 
\begin{align}
dU_{\otimes}=& dU_{\mathsf{A}} + dU_{\mathsf{B}} \nonumber\\
=& \mathrm{Tr}\big[d\varrho_{\mathsf{A}} \hat{H}_{\mathsf{A}}\big]+\mathrm{Tr}\big[d\varrho_{\mathsf{B}} \hat{H}_{\mathsf{B}}\big] \nonumber\\
=&\mathrm{Tr}_{\mathsf{A}}\big[\big(-i[\hat{H}_{\mathsf{A}},\varrho_{\mathsf{A}}]-i \mathrm{Tr}_{\mathsf{B}}[V,\chi]+\mathpzc{L}_{\mathsf{A}}[\varrho_{\mathsf{A}}]\big) \hat{H}_{\mathsf{A}}\big]
+
\mathrm{Tr}_{\mathsf{B}}\big[\big(-i[\hat{H}_{\mathsf{B}},\varrho_{\mathsf{B}}]-i \mathrm{Tr}_{\mathsf{A}}[V,\chi]+\mathpzc{L}_{\mathsf{B}}[\varrho_{\mathsf{B}}]\big) \hat{H}_{\mathsf{B}}\big]\nonumber
\\
\overset{(\ref{eq:SubsystemLindbladEquations})}{=}&\mathrm{Tr}_{\mathsf{A}}\big[\big(-i \mathrm{Tr}_{\mathsf{B}}[V,\chi]+\mathpzc{L}_{\mathsf{A}}[\varrho_{\mathsf{A}}]\big) \hat{H}_{\mathsf{A}}\big]
+
\mathrm{Tr}_{\mathsf{B}}\big[\big(-i \mathrm{Tr}_{\mathsf{A}}[V,\chi]+\mathpzc{L}_{\mathsf{B}}[\varrho_{\mathsf{B}}]\big) \hat{H}_{\mathsf{B}}\big]\nonumber
\\
=&-i\mathrm{Tr}\big[ [\hat{H}_{\mathsf{A}},V]\chi\big]-i\mathrm{Tr}\big[ [\hat{H}_{\mathsf{B}},V]\chi]\big]
+\mathrm{Tr}_{\mathsf{A}}\big[\mathpzc{L}_{\mathsf{A}}[\varrho_{\mathsf{A}}]\hat{H}_{\mathsf{A}}\big]
+\mathrm{Tr}_{\mathsf{B}}\big[\mathpzc{L}_{\mathsf{B}}[\varrho_{\mathsf{B}}]\hat{H}_{\mathsf{B}}\big]
\nonumber
\\
=&-i\mathrm{Tr}\big[ [\hat{H}_{\mathsf{A}}+\hat{H}_{\mathsf{B}},V]\chi\big]
+\mathrm{Tr}\big[\mathpzc{L}_{\mathsf{A}}[\varrho_{\mathsf{A}}\otimes \varrho_{\mathsf{B}}]\, H_{\mathsf{A}}\big]
+\mathrm{Tr}\big[\mathpzc{L}_{\mathsf{B}}[\varrho_{\mathsf{A}} \otimes\varrho_{\mathsf{B}}]\, H_{\mathsf{B}}\big]
+\mathrm{Tr}_{\mathsf{A}}\big[\mathpzc{L}_{\mathsf{A}}[\varrho_{\mathsf{A}}] \, \mathrm{Tr}_{\mathsf{B}}[V \varrho_{\mathsf{B}}]\big]
+\mathrm{Tr}_{\mathsf{B}}\big[\mathpzc{L}_{\mathsf{B}}[\varrho_{\mathsf{B}}] \, \mathrm{Tr}_{\mathsf{A}}[V \varrho_{\mathsf{A}}]\big]
\nonumber\\
=&-i\mathrm{Tr}\big[ [\hat{H}_{\mathsf{A}}+\hat{H}_{\mathsf{B}},V]\chi\big]
+\mathrm{Tr}\big[\mathpzc{L}_{\mathsf{A}}[\varrho_{\mathsf{A}}\otimes \varrho_{\mathsf{B}}]\, H_{\mathsf{A}}\big]
+\mathrm{Tr}\big[\mathpzc{L}_{\mathsf{B}}[\varrho_{\mathsf{A}} \otimes\varrho_{\mathsf{B}}]\, H_{\mathsf{B}}\big]
+\mathrm{Tr}\big[\mathpzc{L}_{\mathsf{A}}[\varrho_{\mathsf{A}}] \, V \varrho_{\mathsf{B}}\big]
+\mathrm{Tr}\big[\mathpzc{L}_{\mathsf{B}}[\varrho_{\mathsf{B}}] \, V \varrho_{\mathsf{A}}\big]\nonumber
\\
=&-i\mathrm{Tr}\big[ [\hat{H}_{\mathsf{A}}+\hat{H}_{\mathsf{B}},V]\chi\big]
+\mathrm{Tr}\big[\mathpzc{L}_{\mathsf{A}}[\varrho_{\mathsf{A}}\otimes \varrho_{\mathsf{B}}]\, H_{\mathsf{A}}\big]
+\mathrm{Tr}\big[\mathpzc{L}_{\mathsf{B}}[\varrho_{\mathsf{A}} \otimes\varrho_{\mathsf{B}}]\, H_{\mathsf{B}}\big]
+\mathrm{Tr}\big[\mathpzc{L}_{\mathsf{A}}[\varrho_{\mathsf{A}} \otimes \varrho_{\mathsf{B}}] \, V\big]
+\mathrm{Tr}\big[\mathpzc{L}_{\mathsf{B}}[\varrho_{\mathsf{A}}\otimes \varrho_{\mathsf{B}}] \, V \big]\nonumber
\\
=&-i\mathrm{Tr}\big[ [\hat{H}_{\mathsf{A}}+\hat{H}_{\mathsf{B}},V]\chi\big]
+\mathrm{Tr}\big[\mathpzc{L}_{\mathsf{A}}[\varrho_{\mathsf{A}}\otimes \varrho_{\mathsf{B}}]\, (H_{\mathsf{A}}+V)\big]
+\mathrm{Tr}\big[\mathpzc{L}_{\mathsf{B}}[\varrho_{\mathsf{A}} \otimes\varrho_{\mathsf{B}}]\, (H_{\mathsf{B}}+V)\big]\nonumber
\\
=&-i\mathrm{Tr}\big[ [\hat{H}_{\mathsf{A}}+\hat{H}_{\mathsf{B}},V]\chi\big]
+\mathrm{Tr}\big[\mathpzc{L}_{\mathsf{A}}^{\#}[H_{\mathsf{A}}+V]\, \varrho_{\mathsf{A}}\otimes \varrho_{\mathsf{B}}\big]
+\mathrm{Tr}\big[\mathpzc{L}_{\mathsf{B}}^{\#}[H_{\mathsf{B}}+V]\, \varrho_{\mathsf{A}} \otimes\varrho_{\mathsf{B}}\big].
\end{align}

We note that $\mathpzc{L}_{\mathsf{A}}^{\#}[\mathbbmss{I}_{\mathrm{A}}]=\mathpzc{L}_{\mathsf{B}}^{\#}[\mathbbmss{I}_{\mathrm{B}}]= \mathpzc{L}_{\mathsf{A}}^{\#}[\mathbbmss{I}\otimes H_{\mathsf{B}}]=\mathpzc{L}_{\mathsf{B}}^{\#}[H_\mathsf{A} \otimes \mathbbmss{I}] = 0$. 
By adding these vanishing terms to the last line of the above equation, $dU_{\otimes}$ can be rewritten as
\begin{align}
dU_{\otimes}=-i\mathrm{Tr}\big[ [\hat{H}_{\mathsf{A}}+\hat{H}_{\mathsf{B}},V]\chi\big]
+\mathrm{Tr}\big[\big(\mathpzc{L}_{\mathsf{A}}^{\#}[H]+\mathpzc{L}_{\mathsf{B}}^{\#}[H]\big)\, \varrho_{\mathsf{A}}\otimes \varrho_{\mathsf{B}}\big].
\end{align}
According to the above equation, to have a local-energy preserving dynamics, it is sufficient that the following two conditions are satisfied at all times for all states: 
\begin{gather}
[\hat{H}_{\mathsf{A}}+\hat{H}_{\mathsf{B}},V]=0,\\
\mathpzc{L}_{\mathsf{A}}^{\#}[H]+\mathpzc{L}_{\mathsf{B}}^{\#}[H]=0.
\end{gather}
As explained in the main text, the second condition leads also to the total energy conservation, which is in contrast to our objective. 
\twocolumngrid

\end{widetext}

%

\end{document}